\documentclass[prl,aps,epsf,psfig,floats,preprint]{revtex4}
\usepackage{epsfig}
\begin{document}
\def\beq{\begin{equation}}
\def\eeq{\end{equation}}
\def\ber{\begin{eqnarray}}
\def\eer{\end{eqnarray}}
\def\l{\Lambda}
\def\b{{\rm b}}
\def\m{{\rm m}}
\def\om{\Omega_{0\rm m}}
\def\oml{\Omega_\ell}
\def\omx{\Omega_{\l_{\rm b}}}
\def \lleq {\lower0.9ex\hbox{ $\buildrel < \over \sim$} ~}
\def \ggeq {\lower0.9ex\hbox{ $\buildrel > \over \sim$} ~}

\def\apj{{Astrophys.\@ J.\ }}
\def\mn{{Mon.\@ Not.@ Roy.\@ Ast.\@ Soc.\ }}
\def\asta{{Astron.\@ Astrophys.\ }}
\def\aj{{Astron.\@ J.\ }}
\def\prl{{Phys.\@ Rev.\@ Lett.\ }}
\def\prd{{Phys.\@ Rev.\@ D\ }}
\def\pd{{Phys.\@ Rev.\@ D\ }}
\def\nucp{{Nucl.\@ Phys.\ }}
\def\nat{{Nature\ }}
\def\plb {{Phys.\@ Lett.\@ B\ }}
\def \jetpl {JETP Lett.\ }
\def\etal{{\it et al.}}
\def\ie {{\it ie}}
\def\n {\noindent}

\title{Confronting Braneworld Cosmology 
with Supernova data and Baryon Oscillations}

{\author{Ujjaini Alam$^{a,b}$ and Varun Sahni$^a$}
\address{$^a$ Inter-University Centre for Astronomy and Astrophysics,
Post Bag 4, Ganeshkhind\\
Pune 411~007, India}
\address{$^b$International Centre for Theoretical Physics, Strada Costiera 11,
34100 Trieste, Italy}
%Portsmouth~PO1~2EG, Britain}

%\maketitle

\thispagestyle{empty}

\sloppy

\begin{abstract}
\small{
Braneworld cosmology has several attractive and distinctive features.
For instance the effective equation of state in braneworld models can
be both quintessence-like ($w_0 \geq -1$) as well as phantom-like
($w_0 \leq -1$).  Models with $w_0 \geq -1$ ($w_0 \leq -1$) are
referred to as Brane\,2 (Brane\,1) and correspond to complementary
embeddings of the brane in the bulk.  (The equation of state in
Brane\,1 can successfully cross the `phantom divide' at $w = -1$.)
%The acceleration of the universe in a class of braneworld models can
%be a transient feature because of which these models do not contain an
%event horizon and are therefore better able to satisfy the demands of
%string/M-theory.  
In this paper we compare the predictions of braneworld models to two
recently released supernova data sets: the `Gold' data (Riess \etal,
2004) and the data from the Supernova Legacy Survey (SNLS) (Astier
\etal, 2005). We also incorporate the recent discovery of the baryon
acoustic peak in the Sloan Digital Sky Survey (Eisenstein \etal, 2005)
into our analysis.  Our main results are that braneworld models
satisfy both sets of SNe data.  Brane\,1 (with $w_0 \leq -1$) shows
very good agreement with data for values of the matter density bounded
{\em from below}: $\om \ggeq 0.25$ (Gold) and $\om \ggeq 0.2$
(SNLS). On the other hand Brane\,2 (with $w_0 \geq -1$) shows
excellent agreement with data for values of the matter density which
are bounded {\em from above}: $\om \lleq 0.45$ (Gold) and $\om \lleq
0.35$ (SNLS).
%For higher (lower) values of the matter density the Brane\,1 (Brane\,2)
%can provide better agreement to the existing data than $\Lambda$CDM.
The DGP model is excluded at $3\sigma$ by SNLS and at $1\sigma$ by the
Gold dataset. Braneworld models with future `quiescent' singularities
(at which the Hubble parameter and the matter density remain finite
but higher derivatives of the expansion factor diverge) are excluded
by both datasets.  }
\end{abstract}

\maketitle

%\pacs{PACS number(s): 04.50.+h, 98.80.Hw}

\bigskip \medskip
\section{Introduction}
\label{sec:intro}

One of the most remarkable discoveries of the past decade has been the
observation that the expansion of the universe is speeding up rather
than slowing down.  The case for an accelerating universe was first
made on the basis of high redshift type Ia supernovae
\cite{perl98a,perl98b,riess98}, and has since received support from
deeper and better quality SNe data
\cite{tonry03,knop03,barris04,riess04,astier05} as well as
observations of the cosmic microwave background and large scale
structure \cite{spergel03,tegmark03b}.

Theoretically, an accelerating universe can be constructed in a number
of distinct ways (see \cite{ss00} and references therein).  However
three approaches have received considerable attention in the
literature, these are:

\begin{itemize}

\item 
{\em The cosmological constant}. The acceleration of the universe is
caused by the cosmological constant which satisfies $T_{ik} = \Lambda
g_{ik}$ and, hence, has an equation of state $p = -\rho$.  The
combination of a $\Lambda$-term and cold dark matter results in the
$\Lambda$CDM model which appears to provide excellent agreement with
cosmological observations and, when combined with inflationary
predictions of an (almost) scale invariant spectrum of density
perturbations, comprises what may be called the
%$\Lambda$CDM is the leading contender for the privileged
`standard model of cosmology'.

However, despite its long and chequered history (since its inception
by Einstein in 1917), a firm theoretical basis for a small
$\Lambda$-term has so far eluded researchers \cite{wein89}.  Indeed,
the value of the cosmological constant predicted by quantum field
theory is at least $10^{55}$ times larger than its observed value
$\rho_{\rm vac} = \Lambda/8\pi G \simeq 10^{-47}$GeV$^4$, indicated by
recent observations. This fact, taken together with the unevolving
nature of $\Lambda$, suggests that the present epoch may be quite
special since $\Omega_\Lambda \simeq 2\om$. The resulting {\em cosmic
  coincidence} and the high degree of fine tuning associated with a
small $\Lambda$-term have lead physicists to look for alternatives to
the cosmological constant hypothesis.

\item 
{\em Dark Energy}. The expansion of the universe is governed by the
field equations of general relativity (GR), but one (or more)
components of `matter' violate the strong energy condition (SEC) $\rho
+ 3p \geq 0$ thereby causing the universe to accelerate.  To this
category belong quintessence models, the Chaplygin gas, topological
defects and numerous other models of dark energy.  In general, dark
energy can be characterized by an equation of state $w = p/\rho$,
where the observationally determined value of $w$ can be used to
constrain parameters of a given dark energy model.  Although most DE
models have $w \neq {\rm constant}$ some might say that the
cosmological constant with $w = -1$ also belongs to this category.

\item 
{\em Geometric approaches to acceleration}.  The late-time
acceleration of the universe is caused by a departure of space-time
physics from standard GR on large scales and/or at late times. An
important example of this class of models is braneworld cosmology
according to which our three dimensional universe is a lower
dimensional `brane' embedded in a higher dimensional `bulk'
space-time. Braneworld models may provide a low energy manifestation
of string/M-theory \cite{hw,RS}.  Within the cosmological context
braneworld models provide exciting new possibilities some of which are
summarized below (also see \cite{maartens04}).

\noindent
(i) `Quintessential Inflation' \cite{pv99} based on the
Randall-Sundrum (RS) model \cite{RS} may provide a compelling
explanation of both early and late-time acceleration within a single
unified setting \cite{copeland}.  (ii) The Dvali-Gabadadze-Porrati
(DGP) model can lead to an accelerating universe without the presence
of a cosmological constant or some other form of dark energy
\cite{DGP,DDG}.  (iii) A family of braneworld models \cite{ss02} which
unify the approaches of RS and DGP allow the {\em effective} equation
of state of dark energy to be `quintessence-like' $w \geq -1$ as well
as `phantom-like' $w < -1$. In a subclass of these models the
acceleration of the universe is a {\em transient} phenomenon, which
gives way to matter dominated expansion in the future. The absence of
horizons in a transiently accelerating space-time is an attractive
feature of this scenario since it can reconcile current observations
of acceleration with the demands of string/M-theory \cite{horizon}.
(iv) Another aspect of the braneworld cosmology \cite{ss02} is the
possibility of fundamentally new cosmological behaviour ({\em
  loitering} \cite{loiter} \& {\em mimicry} \cite{mimic}) at
moderately high redshifts.  Loitering and mimicry models remain close
to $\Lambda$CDM in the future (hence providing excellent agreement
with SNe data) while departing from $\Lambda$CDM-like behaviour in the
past.
%This flexibility may allow these 
The older age of these braneworld models might make them better
equipped to explain the existence of high redshift QSO's and also
allow for a lower redshift of reionization than that predicted in
$\Lambda$CDM cosmology \cite{richards03,bh,reionization,age}.

\end{itemize}

Whether the increasingly large number of low and high redshift
observations can be accommodated within the braneworld paradigm is an
important subject demanding extensive exploration. Our purpose in this
paper will be more modest, we shall examine the braneworld models
proposed in \cite{ss02} in the light of the Gold SNe data set
\cite{riess04} and the 71 new SNe discovered by the Supernova Legacy
Survey \cite{astier05}. We shall use this data in conjunction with the
recent discovery of the baryon acoustic peak in the Sloan Digital Sky
Survey \cite{eisenstein05} to place constraints on the parameter space
of accelerating braneworld models. An outline of our paper is as
follows: in section II we briefly describe braneworld cosmology, while
section III is devoted to testing braneworld against observations.
Our results and conclusions are presented in section IV.

\section{Accelerating Braneworld Universe }
\label{sec:braneworld}

The equations of motion governing the braneworld can be derived from
the action \cite{CH,Shtanov1}
\beq \label{action}
S = M^3 \left[\int_{\rm bulk} \left( R_5 - 2 \l_{\rm b} \right) - 2
\int_{\rm brane} K \right] + \int_{\rm brane} \left( m^2 R_4 - 2 \sigma \right)
+ \int_{\rm brane} L \left( h_{\alpha\beta}, \phi \right) \, .
\eeq
Here, $R_5$ is the scalar curvature of the metric $g_{ab}$ in the
five-dimensional bulk, and $R_4$ is the scalar curvature of the
induced metric $h_{\alpha\beta}$ on the brane. The quantity $K =
K_{\alpha\beta} h^{\alpha\beta}$ is the trace of the extrinsic
curvature $K_{\alpha\beta}$ on the brane defined with respect to its
{\em inner\/} normal. $L (h_{\alpha\beta}, \phi)$ is the
four-dimensional matter field Lagrangian, $M$ and $m$ denote,
respectively, the five-dimensional and four-dimensional Planck masses,
$\l_{\rm b}$ is the bulk cosmological constant, and $\sigma$ is the
brane tension.  Integrations in (\ref{action}) are performed with
respect to the volume elements on the bulk and brane.

The action (\ref{action}) presents a synthesis of the
higher-dimensional ansatzes proposed by Randall and Sundrum \cite{RS}
and Dvali, Gabadadze, and Porrati \cite{DGP}. An important role in
(\ref{action}) is played by the $m^2\int R_4$ term. This term was
first introduced by Sakharov in a seminal paper \cite{Sakharov} to
describe the backreaction of quantum fluctuations of matter fields
(which, in our case, reside on the brane). Its presence is crucial in
making the braneworld accelerate since it
%The presence of the brane curvature term $m^2\int_{\rm brane}R_4$ in
%(\ref{action}) 
introduces an important length scale 
$$\ell = 2m^2/M^3~.$$
On short length scales $r \ll \ell$ (early times) one
recovers general relativity, whereas on large length scales $r \gg \ell$
(late times) brane-specific effects begin to play an important role,
leading to the acceleration of the universe \cite{DGP,ss02}.

The cosmological evolution of a spatially flat braneworld is described
by the Hubble parameter
\begin{equation}\label{eq:hubble}
H^2 (a)  = {A \over a^3} + B + {2 \over \ell^2} \left[ 1 \pm \sqrt{1 + \ell^2
\left({A \over a^3} + B  - {\Lambda_{\rm b} \over 6} - {C \over a^4} \right)}
\right] \, , 
%\equiv f(a)  \, , 
\end{equation}
where
\begin{equation}\label{eq:bao}
A = {\rho_{0} a_0^3 \over 3 m^2} \, , \quad B = {\sigma \over 3 m^2} \, , 
\end{equation}
and the `dark radiation' term $C/a^4$ describes the projection of bulk
degrees of freedom onto the brane.  (Note that the four-dimensional
Planck mass $m$ is related to the effective Newton's constant on the
brane as $m = 1/\sqrt{8\pi G}$.)

The two signs in (\ref{eq:hubble}) correspond to the two distinct ways in
which the brane can be embedded in the higher dimensional bulk.
%The braneworld with the
%lower/upper sign in (\ref{eq:hubble}) expands
%faster/slower than $\Lambda$CDM. 
%(The underlined term in (\ref{eq:hubble}) makes braneworld dynamics different
%from standard FRW cosmology.) 
Three limiting cases of our model may be of interest
to the reader: 

\begin{enumerate}

\item $m=0$ in (\ref{eq:hubble}) corresponds to the well known
FRW generalization of the RS scenario
\beq\label{eq:cosmolim}
H^2 + {\kappa \over a^2} = {\l_{\rm b} \over 6} + {C \over a^4} + {(\rho +
\sigma)^2 \over 9 M^6}\, .
\eeq

\item $M=0$ in (\ref{eq:hubble}) gives rise to $\Lambda$CDM 
\beq
H^2 (a)  = {A \over a^3} + B \,.
\eeq

\item Finally, by setting $\sigma = 0$ and $\l_{\rm b} = 0$ in (\ref{eq:hubble})
we recover the DGP model \cite{DGP}.

The Braneworld models proposed in \cite{ss02} fall into two main
categories:

\end{enumerate}

\begin{itemize}
\item
{{\bf Brane\,1} (B1):} The lower sign in (\ref{eq:hubble}) leads to
the following form of the Hubble parameter \footnote{We have neglected
  the dark radiation term for simplicity.}:
\beq
H^2 (a)  = {A \over a^3} + \Lambda_{\rm eff} \, .
\eeq
The effective cosmological `constant' $\Lambda_{\rm eff}$ is composed
of two terms: a constant $\Lambda$-term and a `screening term'
\cite{lue-starkman}:
\beq\label{eq:screening}
\Lambda_{\rm eff} = \underbrace{ (B + {2 \over \ell^2}\large ) }
-  \underbrace{{2 \over \ell^2}\sqrt{1 + \ell^2
\left({A \over a^3} + B - \frac{\l_b}{6}\right)  }}
 \, \nonumber
 \eeq
 \beq
 \hskip -0.5cm \Downarrow \hskip 3.3cm \Downarrow
 \eeq
 \hskip 5.5cm {\Large{$\Lambda$}} \hskip 1.9cm Screening term

\bigskip
Since the screening term decreases with time, the value of {\em the
effective cosmological constant $\Lambda_{\rm eff}$ increases}. In
this respect Brane\,1 resembles phantom cosmology which has $w < -1$.
It is important to note, however, that in the braneworld case there is
{\em no violation of the weak energy condition and also no future `big
rip' singularity}. Indeed, from (\ref{eq:screening}) it is quite clear
that the universe evolves to $\Lambda$CDM in the future.

Since our main desire in this paper will be to test braneworld models
against observations, it will be helpful to recast
Eq. (\ref{eq:hubble}) with the lower sign in the following form
\beq \label{eq:hubble1}
{H^2(z) \over H_0^2} = \om (1\!+\!z)^3 + 
\Omega_\sigma + \underline{2 \oml - 2 \sqrt{\oml}\,
\sqrt{\om (1\!+\!z )^3 + \Omega_\sigma + \oml +
\omx}} \, ,
\eeq
where $z = a_0/a(t)-1$ is the cosmological redshift, while
\beq \label{eq:omegas}
\om =  {\rho_0 \over 3 m^2 H_0^2} \, , 
\quad \Omega_\sigma = {\sigma \over 3 m^2
H_0^2} \, , \quad \oml = {1 \over l^2 H_0^2} \, , \quad
\omx = - {\l_{\rm b} \over 6 H_0^2} \, ,
\eeq
are dimensionless parameters whose values must be determined from
observations.  $\Omega_\sigma$ is determined by the constraint
relation
\beq \label{eq:omega1}
\om + \Omega_\sigma - \underline{2
\sqrt{\oml}\, \sqrt{1 + \omx}} = 1.
\eeq
{\em  The difference between $\Lambda$CDM and braneworld
cosmology is brought about by the underlined terms in
(\ref{eq:hubble1}) \& (\ref{eq:omega1}).}

\item{{\bf Brane\,2} (B2):} The upper sign in (\ref{eq:hubble}) results in
\beq \label{eq:hubble2}
{H^2(z) \over H_0^2} = \om (1\!+\!z)^3 + 
\Omega_\sigma + \underline{2 \oml + 2 \sqrt{\oml}\,
\sqrt{\om (1\!+\!z )^3 + \Omega_\sigma + \oml +
\omx}} \, ,
\eeq
where $\oml < 1 + \omx$
and $\Omega_\sigma$ is determined from
\beq \label{eq:omega2}
\om + \Omega_\sigma + \underline{2
\sqrt{\oml} \, \sqrt{1 + \omx}} = 1.
\eeq

{\em  The difference between $\Lambda$CDM and braneworld
cosmology is brought about by the underlined terms in 
(\ref{eq:hubble2})
\& (\ref{eq:omega2}).}

The two models Brane\,1 and Brane\,2 are complementary and reflect the
two distinct ways in which the brane can be embedded in the bulk.

From (\ref{eq:hubble1}) \& (\ref{eq:hubble2}) it is easy to see that
both braneworld models approach the standard matter dominated universe
at early times [with a small correction term $\sim (1+z)^{3/2}$]. At
late times the behaviour of the braneworld can differ from both
$\Lambda$CDM and SCDM.  This feature makes braneworld models testable
and allows the braneworld scenario to provide a new explanation for
the observational discovery of an accelerating universe.

The expansion of the braneworld can be characterized by the
deceleration parameter
\beq \label{eq:decel}
q(z) = \frac{H'(z)}{H(z)} (1+z) - 1 \, ,
\eeq
and the {\em effective} equation of state
\beq \label{eq:state0}
w(z) = {2 q(z) - 1 \over 3 \left[ 1 - \Omega_{\rm m}(z) \right] } \, .
\eeq
From (\ref{eq:hubble1}), (\ref{eq:hubble2}), (\ref{eq:decel}) and
(\ref{eq:state0}) it is easy to obtain the following expression for
the current value of the effective equation of state \cite{ss02}
%\begin{equation}
\beq
%w_0 %= {2 q_0 - 1 \over 3 \left( 1 - \om \right)} ,\nonumber\\
w_0 = - 1 \pm
{\om \over 1 - \om} \, {\sqrt{\oml \over
\om + \Omega_\sigma + \oml + \omx}} \, ,
%&=& - 1 \pm {\om \over 1 - \om} \,
%\frac{\sqrt{\oml}}{\sqrt{1+\omx} \mp \sqrt{\oml}}~.
\label{eq:brane_state}
\eeq
%\end{equation}
and we immediately find that $w_0 \leq -1$ when we take the lower sign
in (\ref{eq:brane_state}), which corresponds\footnote{ It is
  interesting to note that Brane\,1 models have $w_0 \leq -1$ and
  $w(z) \simeq -0.5$ at $z \gg 1$.  They therefore successfully cross
  the `phantom divide' at $w=-1$.}  to Brane\,1.  The second choice
(Brane\,2) gives $w_0 \geq -1$.

\item {\em Mimicry models}. It is interesting to note that 
for values of $z$ and $\Omega_{\Lambda_{\rm b}}$
\& $\Omega_\ell$ satisfying
\beq
\om (1\!+\!z)^3 \ll \left( \sqrt{1+\omx} \mp
\sqrt{\oml} \right)^2 \, , \label{eq:mimic0}
\eeq
Eqs.~(\ref{eq:hubble1}) and (\ref{eq:hubble2}) reduce to
\beq
{H^2(z) \over H_0^2} \simeq \Omega^{\rm \Lambda{\rm CDM}}_{\rm m} (1\!+\!z)^3 + 1 -
\Omega^{\rm \Lambda{\rm CDM}}_{\rm m}\, ,
\label{eq:lcdm}
\eeq
where the new density parameter $\Omega^{\rm \Lambda{\rm CDM}}_{\rm m}$ is defined by the relation
\beq \label{lcdm}
\Omega^{\rm \Lambda{\rm CDM}}_{\rm m} = {\alpha \over \alpha \mp 1}\, \Omega_{\rm m}  \, ,~
~ 
%\begin{equation} \label{alpha} 
\alpha = {\sqrt{1 + \omx} \over \sqrt{\oml}} \, .
%\end{equation}
\eeq

The braneworld therefore displays a remarkable property called {\em
``cosmic mimicry''}: at low redshifts, the Brane\,1 universe expands
as $\Lambda$CDM (\ref{eq:lcdm}) with $\Omega^{\rm \Lambda{\rm
CDM}}_{\rm m} < \Omega_{\rm m}$ [$\Omega^{\rm \Lambda{\rm CDM}}_{\rm
m}$ is determined by (\ref{lcdm}) with the lower (``$+$'') sign].  The
Brane\,2 model at low redshifts also mimics $\Lambda$CDM but with a
{\em larger value\/} of the density parameter $\Omega^{\rm \Lambda{\rm
CDM}}_{\rm m} > \Omega_{\rm m}$ with $\Omega^{\rm \Lambda{\rm
CDM}}_{\rm m}$ being determined by (\ref{lcdm}) with the upper
(``$-$'') sign.

The range of redshifts over which cosmic mimicry occurs is given by $0
\leq z \ll z_{\m}$, where
\beq
z_{\m} = \frac{ \left(\sqrt{1 + \omx} \mp
\sqrt{\oml} \right)^{2/3}} {\om^{1/3}} - 1 \, .
\label{eq:mimic1}
\eeq

As demonstrated in \cite{mimic} the Hubble parameter in mimicry models
departs from that in $\Lambda$CDM at {\em intermediate\/} redshifts
($z > z_{\rm m} \sim ~{\rm few}$). This could lead to interesting
cosmological effects since the age of the high redshift universe in a
mimicry model can be greater than that in $\Lambda$CDM while the
redshift of reionization can be lower \cite{mimic}.  Since the mimicry
models and $\Lambda$CDM are virtually indistinguishable at lower
redshifts, both are expected to fit the SNe data (at $z < 2$) equally
well.

\end{itemize}

\section{Comparing braneworld models with observations}
\label{sec:comparison}

In this paper we shall compare the braneworld model \cite{ss02}
against three sets of observations. We briefly summarize each of the
data sets which we shall use before proceeding to give the results of
our comparison.

\begin{enumerate}
  
\item 
{\em The Gold SNe data set} : As recently as 2003, the entire
supernova dataset from the two different surveys -- Supernova
Cosmology Project (SCP) and High $z$ Supernova Search Team (HZT),
along with low redshift supernovae from Calan-Tololo Supernova Search
(CTSS) comprised of a meager $92$ supernovae
\cite{perl98a,riess98,perl98a}, with very few at high redshifts,
$z>0.7$. The method of data reduction for the different teams was also
somewhat different, so that it was not possible to use the supernovae
from the two datasets concurrently.  The picture changed somewhat
dramatically during 2003-2004, when a set of papers from both these
teams \cite{tonry03,knop03,barris04} presented a joint dataset of
$194$ supernovae which used the same data reduction method. This new
data resulted in doubling the dataset at $z>0.7$.  Not all these
supernovae could be identified beyond doubt as Type Ia supernovae
however, in many cases complete spectral data was not available. In
early 2004, Riess \etal \cite{riess04} reanalyzed the data with
somewhat more rigorous standards, excluding several supernovae for
uncertain classification or inaccurate colour measurements. They also
added $14$ new high redshift supernovae observed by the Hubble Space
Telescope (HST) to this sample. The resultant sample comprises of
$157$ supernovae (the furthest being at redshift $z=1.75$) which have
been classified as Type Ia supernovae beyond doubt-- the `Gold'
dataset. We shall be using this `Gold' dataset as our first supernova
sample.

\item
{\em The Supernova Legacy Survey SNe data set (SNLS)} : The SuperNova Legacy
Survey \cite{snls} is an ongoing 5-year project which is expected to
yield more than $700$ spectroscopically confirmed supernovae below
redshift of one. The first year results from this survey
\cite{astier05} have provided us with 71 new supernovae below $z=1$.
We shall use these 71 SNe together with the already available low-$z$
supernova data, \ie \ a total of $115$ SNe, as our second supernova
sample.
  
\item 
{\em The Baryon Acoustic Oscillation Peak (BAO)} : A remarkable confirmation
of the standard big bang cosmology has been the recent detection of a
peak in the correlation function of luminous red galaxies in the Sloan
Digital Sky Survey \cite{eisenstein05}.  This peak, which is predicted
to arise precisely at the measured scale of $100$ h$^{-1}$ Mpc due to
acoustic oscillations in the primordial baryon-photon plasma prior to
recombination, can provide a `standard ruler' with which to test dark
energy models. Specifically, we shall use the value
\cite{eisenstein05}
\beq
A = \frac{\sqrt{\om}}{h(z_1)^{1/3}}~\bigg\lbrack ~\frac{1}{z_1}~\int_0^{z_1}\frac{dz}{h(z)}
~\bigg\rbrack^{2/3} = ~0.469 \pm 0.017~,
\eeq
where $h(z) = H(z)/H_0$ is defined in (\ref{eq:hubble1}) and
(\ref{eq:hubble2}) for Brane\,1 and Brane\,2 respectively, and $z_1 =
0.35$ is the redshift at which the acoustic scale has been measured.

\end{enumerate}

\subsection{Methodology and Results}
\label{sec:results}

For the supernova data, we shall use the $\chi^2$ minimization where 
\beq
\chi^2(H_0,\om,p_j)=\sum_i \frac{[y_{{\rm fit},i}(z_i;H_0,\om,p_j)-y_{i}]^2}{\sigma^2_i}\,\,.
\eeq
Here, $y_i$ is the data at redshift of $z_i$ and $\sigma_i$ is the
uncertainty in the individual $y_i$, and $p_j$ are the braneworld
parameters ($\oml,\omx$ for Brane\,1 and Brane\,2). For the `Gold'
dataset, $y_i=\mu_{0,i}=m_B-M=5{\rm log}d_L+25$, the extinction
corrected distance modulus for SNe at redshift $z_i$. The error
$\sigma(\mu_{0,i})$ includes the uncertainty in galaxy redshifts due
to a peculiar velocity of $400$ km/s. For SNLS,
$y_i=\mu_{B,i}=m_B-M+\alpha(s-1)-\beta c=5\log_{10} (d_L/10 {\rm
  pc})$.  The error $\sigma(\mu_{B,i})$ includes effects due to a
peculiar velocity of $300$ km/s. We assume a flat universe for our analysis.

We note that, for the SNLS data, we have to deal with two additional
parameters $\alpha, \beta$ during the minimization. However, the error
$\sigma$ is dependent on $\alpha, \beta$. Therefore, if we minimize
with respect to all the parameters, the process will be biased towards
increasing the errors in order to decrease the $\chi^2$. To avoid
this, we refrain from varying $\alpha, \beta$ together with the
cosmological parameters. At each iteration $\alpha, \beta$ are fixed,
while the remaining parameters are varied to obtain the minimum
$\chi^2$, then the values of $\alpha, \beta$ are changed for the next
iteration. This process is continued till the global minimum is
obtained. This method is equivalent to the method followed by the SNLS
team, and for flat $\l$CDM, our results concur with those in
\cite{snls}. In presenting the final results, the nuisance parameters
$\alpha, \beta, M$ (where $M$ depends on $H_0$) are marginalized over
to obtain bounds on the cosmological parameters of interest.

For Brane\,1 and Brane\,2, the cosmological parameters to be estimated
are $\om, \ \oml \ {\rm and} \ \omx$ ($\Omega_{\sigma}$ is calculated
from Eqs (\ref{eq:omega1}) and (\ref{eq:omega2}) respectively for
Brane\,1 \& Brane\,2). After marginalizing over the statistical
nuisance parameters, we obtain the three-dimensional probability
distribution in the $(\om, \oml, \omx)$ space: $P(\om, \oml, \omx)$.
We perform maximum likelihood analysis on the system with the priors
$0 \leq \om \leq 1, \ \oml \geq 0, \ \omx \geq 0$. For Brane\,2, we
use the added constraint $\oml \leq 1+\omx$.  For Brane\,1 \& Brane\,2
the constraint relations (\ref{eq:omega1}), (\ref{eq:omega2}) combined
with $\Omega_{\kappa}=0$, set the lower bound $\Omega_{\l_b} \geq -1$.
However since $\omx \geq 0$ is a more physically appealing model (it
includes anti-de Sitter space (AdS) bulk geometry), we choose this as
a prior for further analysis.

We may add further information to the analysis from the baryon
acoustic oscillation (BAO) data. We obtain the joint probability
distribution for the SNe data and the BAO data as $P(\chi^2_{\rm
SNe+BAO})=P(\chi^2_{\rm SNe}) P(A)$, where $A$ is the quantity
defined in eq~(\ref{eq:bao}) and we assume that it follows a Gaussian
probability distribution with mean $\bar{A}=0.469$ and an error of
$\sigma=0.017$.

In the figure~\ref{fig:chi}, we show the results for Brane\,1 and Brane\,2
using both Gold and SNLS data, in conjunction with the baryon
acoustic oscillation peak (BAO). We show the reduced $\chi^2$ per
degrees of freedom as a function of $\om$, marginalized over $\oml,
\omx, H_0$. For Brane\,1, we find that the supernova data alone, in both
cases, favours a somewhat larger value of $\om$ at the minimum, with
`Gold' preferring a higher value than SNLS. When used in conjunction
with BAO, however, both datasets prefer a matter density of $\om
\simeq 0.26$. For Brane\,2, the preferred value of $\om$ is around
$\om=0.2$ for SNLS and around $\om=0.3$ for `Gold'. When used with
BAO, once again, $\om \simeq 0.26$ is preferred.

During the analysis we find that for both Brane\,1 and Brane\,2, the
results are very weakly dependent on the bulk cosmological constant
$\omx$, and marginalizing over $\omx$ does not affect the results very
much. In further analysis, we therefore fix the value of $\omx$ to its
best-fit value of $\omx=0$. 

\begin{figure*}
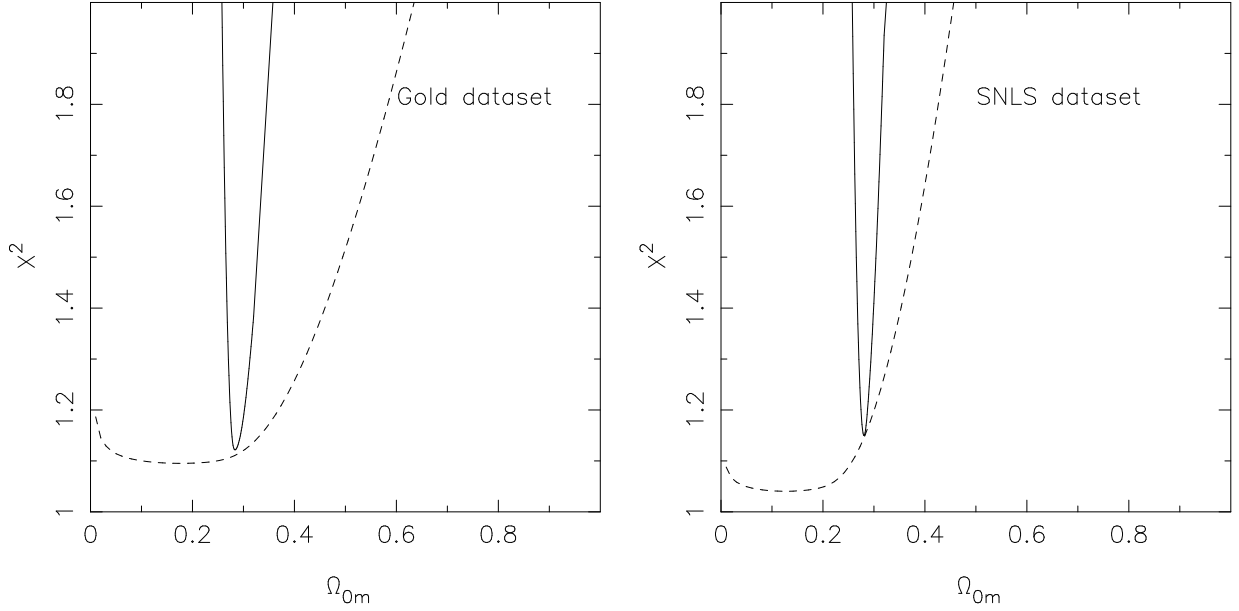
 
\centering
\begin{center}
\vspace{-0.05in}
\centerline{\mbox{Brane\,1}}
\vspace{-0.05in}
%{\mbox{{\hspace{1.0in} $\om=0.3$ \hspace{3.0in} $\om=0.4$}}}
$\begin{array}{c@{\hspace{0.2in}}c}
\multicolumn{1}{l}{\mbox{}} &
\multicolumn{1}{l}{\mbox{}} \\ [-0.20in]
\epsfxsize=3.1in
\epsffile{brane1_riess_chi.epsi} &  
\epsfxsize=3.1in
\epsffile{brane1_snls_chi.epsi} \\
\end{array}$
{\mbox{Brane\,2}}
%\vspace{-0.05in}
%{\mbox{{\hspace{1.0in} $\om=0.3$ \hspace{3.0in} $\om=0.4$}}}
$\begin{array}{c@{\hspace{0.2in}}c}
\multicolumn{1}{l}{\mbox{}} &
\multicolumn{1}{l}{\mbox{}} \\ [-0.15in]
\epsfxsize=3.1in
\epsffile{brane2_riess_chi.epsi} &  
\epsfxsize=3.1in
\epsffile{brane2_snls_chi.epsi} \\  
\end{array}$
\end{center}
%\vspace{-0.27in}
\caption{\small 
The reduced $\chi^2$ per degree of freedom as a function of $\om$,
marginalized over $\oml, \omx, H_0$ for Brane\,1 (top) and Brane\,2
(bottom).  The left panel shows results for the Gold Supernova data
while the right panel shows results for the SNLS data. The dashed line
is each panel is obtained by using supernova data alone, while the
solid line uses both SNe data and the baryon oscillation peak. }
\label{fig:chi}
\end{figure*}

\begin{table}
\caption{\footnotesize Best-fit $\om-\oml$ with corresponding
  $1\sigma$ errors for the Brane\,1 model, with the present value of
  the equation of state $w_0$ and the best-fit $\chi^2$ for the two
  supernova datasets. The best-fit $\chi^2$ for $\l$CDM is also shown
  for comparison.}
\begin{tabular}{c|c|c|c|c|c}
&$\om$&$\oml$&$w_0$&$\chi^2_{\rm min}$&$\chi^2_{\l \rm CDM}$\\ \hline
Gold+BAO&$0.274^{+0.039}_{-0.042}$&$0.0^{+0.031}_{}$&$-1.0^{}_{-0.057}$&$1.13$&$1.13$\\
SNLS+BAO&$0.269^{+0.041}_{-0.034}$&$0.054^{+0.051}_{-0.054}$&$-1.073^{+0.073}_{-0.035}$&$1.08$&$1.15$
\end{tabular}\label{tab:chi_B1}
\end{table}

\begin{figure*}
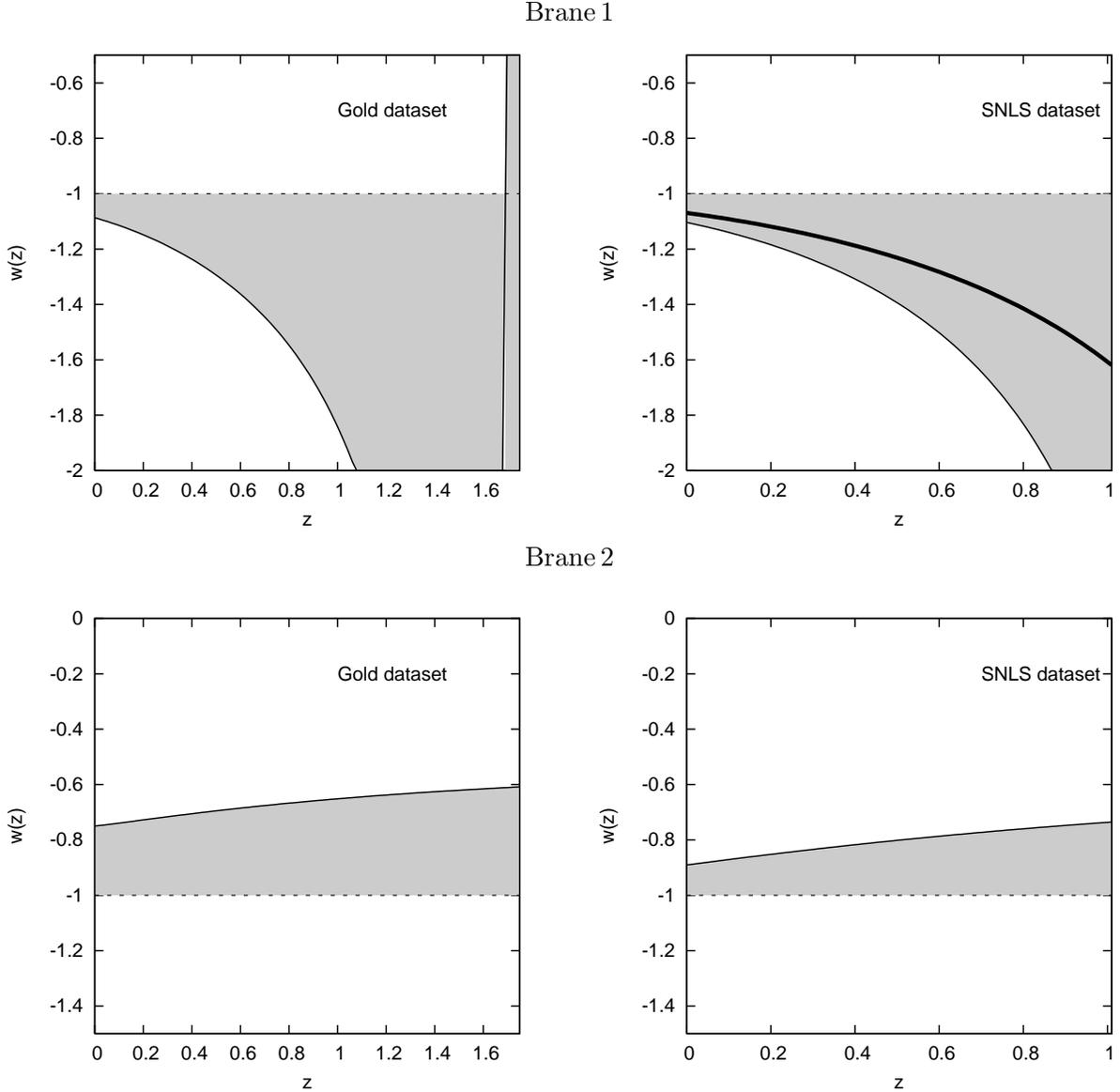
 
\centering
\begin{center}
\centerline{\mbox{Brane\,1}}
\vspace{-0.35in}
$\begin{array}{c@{\hspace{0.2in}}c}
\multicolumn{1}{l}{\mbox{}} &
\multicolumn{1}{l}{\mbox{}} \\ [-0.20in]
\epsfxsize=3.1in
\epsffile{wplot_gold_B1.epsi} &  
\epsfxsize=3.1in
\epsffile{wplot_snls_B1.epsi} \\
\end{array}$
{\mbox{Brane\,2}}
$\begin{array}{c@{\hspace{0.2in}}c}
\multicolumn{1}{l}{\mbox{}} &
\multicolumn{1}{l}{\mbox{}} \\ [-0.15in]
\epsfxsize=3.1in
\epsffile{wplot_gold_B2.epsi} &  
\epsfxsize=3.1in
\epsffile{wplot_snls_B2.epsi} \\  
\end{array}$
\end{center}
\caption{\small
The redshift variation of the ``effective'' equation of state $w(z)$
for the Brane\,1 (top panel) and Brane\,2 (bottom panel) models using
the Gold+BAO (left panel) and SNLS+BAO (right panel) datasets. The
light grey contours denote the $1\sigma$ errors around the best-fit,
and the dashed line represents $\l$CDM, which is the upper (lower)
limit for $w(z)$ for the Brane\,1 (Brane\,2) model. In the top right
panel, the thick solid line represents the best-fit for the SNLS+BAO
data for the Brane\,1 model. For the other three cases, the best-fit
is at the $\l$CDM line. We see that the behaviour of the ``effective''
equation of state of the braneworld models can be markedly different
from $\l$CDM within $1\sigma$ even for the small values of $\oml$
allowed by the data, especially for Brane\,1. For the Gold data, which
extends to higher redshifts, we can even see the existence of a pole
in the ``effective'' equation of state for the Brane\,1 models.}
\label{fig:wplot}
\end{figure*}

\begin{table}
\caption{\footnotesize Best-fit $\om-\oml$ with corresponding
  $1\sigma$ errors for the Brane\,2 model, with the present value of
  the equation of state $w_0$ and the best-fit $\chi^2$ for the two
  supernova datasets. The best-fit $\chi^2$ for $\l$CDM is also shown
  for comparison.}
\begin{tabular}{c|c|c|c|c|c}
&$\om$&$\oml$&$w_0$&$\chi^2_{\rm min}$&$\chi^2_{\l \rm CDM}$\\ \hline
Gold+BAO&$0.277^{+0.051}_{-0.038}$&$0.0^{+0.095}_{}$&$-1.0^{+0.090}_{}$&$1.13$&$1.13$\\
SNLS+BAO&$0.285^{+0.037}_{-0.036}$&$0.0^{+0.043}_{}$&$-1.0^{+0.068}_{}$&$1.15$&$1.15$
\end{tabular}\label{tab:chi_B2}
\end{table}

In the table~\ref{tab:chi_B1} we show the best-fit $\om-\oml$ with
$1\sigma$ errors around it for the Brane\,1 model for the two
supernova datasets, with the $\chi^2$ at the best-fit. The best-fit
$\chi^2$ for $\l$CDM is also shown for comparison. We know that the
supernova data alone is not able to place strong constraints on the
value of the matter density $\om$ for Brane\,1 models for both
datasets (see upper panel of figure~\ref{fig:conf}). If we use the
baryon acoustic oscillation peak in conjunction with the supernova
data, then there are stringent and realistic constraints on the value
of $\om$, therefore we show the $\chi^2$ for the joint probability
distribution only. For the Brane\,1 model, $\oml$ cannot be negative,
and $w_0 \leq -1$ always, and we show the bounds on $\oml$ and $w_0$
accordingly.  For the Gold dataset, the supernova and BAO data in
conjunction favour a $\l$CDM universe with
$\om=0.274^{+0.039}_{-0.042}$ and $\oml=0.0^{+0.031}_{}$. The
``effective'' equation of state at present is
$w_0=-1.0^{}_{-0.057}$. For the SNLS dataset, a braneworld model with
a small value of $\oml$ is slightly favoured over $\l$CDM, with
$\om=0.269^{+0.041}_{-0.034}$ and $\oml=0.054^{+0.051}_{-0.054}$. The
present value of the ``effective'' equation of state is
$w_0=-1.073^{+0.073}_{-0.035}$.  Since $\oml=0$ represents $\l$CDM,
this implies that BAO and SNe data together choose a best-fit Brane\,1
model that is either equivalent to, or very close to, the $\l$CDM
model. However, within $1\sigma$, the value of $\oml$ allowed is
sufficiently large to allow for interesting braneworld behavior,
especially for the SNLS data. 

In the table~\ref{tab:chi_B2}, we similarly show the best-fit
$\om-\oml$ with $1\sigma$ errors around it for the Brane\,2 model for
the two supernova datasets, with the $\chi^2$ at the best-fit, along
with the best-fit for $\l$CDM. As before, since for the Brane\,2
model, $\oml$ cannot be negative, and $w_0 \geq -1$ always, we show
the bounds on $\oml$ and $w_0$ accordingly. In this case, the current
data appears to favour a $\Lambda$CDM universe over the Brane\,2 model
for both datasets. Somewhat larger values of $\oml$ are allowed for
Brane\,2 within $1\sigma$ for the Gold data than for the SNLS data.

The low values of $\oml$ allowed by the current data for both the
braneworld models can still give rise to sufficiently interesting
behaviour. To demonstrate this, in figure~\ref{fig:wplot} we show the
$1\sigma$ confidence levels for the ``effective'' equation of state
$w$ for the Brane\,1 and Brane\,2 models using Gold+BAO and SNLS+BAO
data. We see that, even for the small values of $\oml$ allowed by the
data, the ``effective'' equation of state looks quite different from
the $w=-1$ cosmological constant model within $1\sigma$ especially for
the Brane\,1 model. In fact, for the Gold data, which extends to
higher redshifts, we can see the existence of a singularity in the
``effective'' equation of state at $z\simeq 1.6$ for the Brane\,1
model. We would like to emphasize here that the presence of such poles
in the ``effective'' quantity $w(z)$ does not signal to any inherent
pathologies of the braneworld models described here, since the scale
factor and its derivatives remain well behaved throughout the evolution
of the universe. 
Indeed, the reason for the occurance of a pole is simple and has to do
with the fact that the density parameter $\Omega_m(z)$,
which increases with increasing redshift, crosses unity 
at high $z$ \cite{ss02,loiter}. This results in a pole in the effective
equation of state, since $w(z)$ in (\ref{eq:state0}) 
diverges when $\Omega_m(z) \to 1$.
It should be stressed that the diverging
equation of state is a signature of these braneworld
models. Although it is difficult to reconstruct a diverging
equation of state using standard parametrizations of dark energy, the
equation of state of such braneworld models can be successfully
reconstructed using other reconstruction techniques such as smoothing of
the supernova data \cite{smooth}. So, it is hoped that with better quality
data, it should be possible to test 
braneworld models by studying the behaviour
of the effective equation
of state at high $z$. At present we see that, within $1\sigma$, braneworld models
that are distinctly different from the cosmological constant are able
to satisfy the current supernova data and therefore remain a possible
candidate for dark energy along with the cosmological constant.

\begin{figure*}
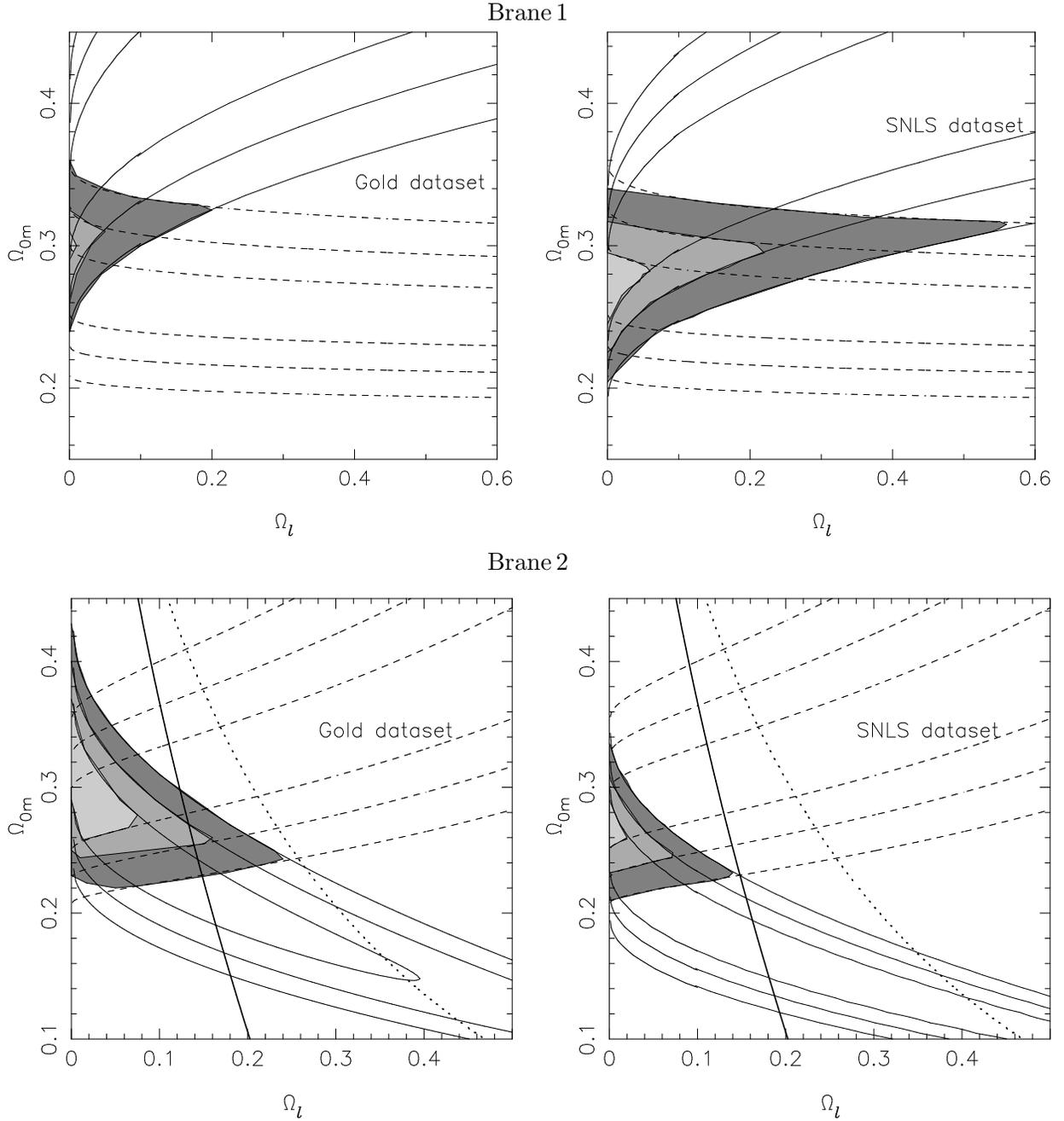
 
\centering
\begin{center}
\vspace{-0.65in}
\centerline{\mbox{Brane\,1}}
\vspace{-0.35in}
%{\mbox{{\hspace{1.0in} $\om=0.3$ \hspace{3.0in} $\om=0.4$}}}
$\begin{array}{c@{\hspace{0.2in}}c}
\multicolumn{1}{l}{\mbox{}} &
\multicolumn{1}{l}{\mbox{}} \\ [-0.20in]
\epsfxsize=3.1in
\epsffile{brane1_riess.epsi} &  
\epsfxsize=3.1in
\epsffile{brane1.epsi} \\
\end{array}$
{\mbox{Brane\,2}}
%\vspace{-0.05in}
%{\mbox{{\hspace{1.0in} $\om=0.3$ \hspace{3.0in} $\om=0.4$}}}
$\begin{array}{c@{\hspace{0.2in}}c}
\multicolumn{1}{l}{\mbox{}} &
\multicolumn{1}{l}{\mbox{}} \\ [-0.15in]
\epsfxsize=3.1in
\epsffile{brane2_riess.epsi} &  
\epsfxsize=3.1in
\epsffile{brane2.epsi} \\  
\end{array}$
\end{center}
\vspace{-0.27in}
\caption{\small 
Confidence levels in the $\om-\oml$ plane for Brane\,1 (top) and
Brane\,2 (bottom).  $H_0$ is marginalized over and $\omx$ is fixed at
the best fit value of $\omx=0$.  The left panel shows results for the
Gold Supernova data while the right panel shows results for the SNLS
data. The solid lines in each panel represent the $1, 2, 3 \sigma$
confidence levels obtained by using supernova data alone, while the
dashed lines are the $1, 2, 3 \sigma$ contours for the baryon
oscillation peak. The light grey, medium grey and dark grey contours
represent the $1, 2, 3 \sigma$ confidence levels when the supernova
data is used in conjunction with the baryon oscillation peak. The
region to the right of the thick dotted line in the lower panels
represents Brane\,2 models which will undergo a future `quiescent'
singularity. The thick solid line in the lower panels represents the
DGP model.}
\label{fig:conf}
\end{figure*}

In figure~\ref{fig:conf}, we further explore the $\om-\oml$ parameter
space for the Brane\,1 and Brane\,2 models. We see from the
$1\sigma,2\sigma,3\sigma$ contours that Brane\,1 satisfies the SNLS
dataset for a larger region in the $\om-\oml$ parameter space as
compared to the Gold dataset. BAO does not depend very strongly on the
value of $\oml$, rather, it is sensitive to the value of $\om$.
Therefore, in conjunction with BAO, Brane\,1 provides a better fit to
the SNLS data than to the Gold data. The situation is just the
opposite in the case of Brane\,2, which provides a better fit to the
Gold data as compared to the SNLS data. 

An interesting feature of braneworld models is that for a finite
region of parameter space, the Brane\,2 universe can expand towards a
`quiescent' future singularity at which the energy density and the
Hubble parameter remain well behaved, but higher derivatives of the
expansion factor ($\stackrel{..}{a}$, $\stackrel{...}{a}$ etc.)
diverge \cite{ss02b}.  From (\ref{eq:hubble2}) we see that a
braneworld model which satisfies
\beq\label{eq:sing}
\Omega_{\sigma}+\oml+\omx < 0.
\eeq
will run into a future singularity at the redshift
\beq\label{eq:sing_time}
z_s=\left( - \frac{\Omega_{\sigma}+\oml+\omx}{\om}
\right)^{1/3}-1 \,\,.
\eeq

The time of occurrence of the singularity (measured from the present
moment) can easily be determined from
\beq
T_s = t(z=z_s) - t(z=0) = \int_{z_s}^0\frac{dz}{(1+z)H(z)},
\eeq
where $H(z)$ is given by (\ref{eq:hubble2}).

From the lower panel of figure~\ref{fig:conf}, we see that universes
which terminates in a `quiescent' future singularity are excluded at
the $3\sigma$ confidence level for both SNLS and Gold datasets when
used in conjunction with the baryon oscillations.

As noted in section~\ref{sec:braneworld}, the DGP model forms a
special case of our braneworld cosmology obtained by putting
$\omx=\Omega_{\sigma}=0$ in equation~\ref{eq:hubble2}.  We see that
using the SNLS data together with BAO, we may narrowly rule out the
DGP model of braneworld dark energy at $3\sigma$ (thick solid line in
right lower panel of figure~\ref{fig:conf}). However, for the Gold
data and BAO, the flat DGP model is acceptable within $2\sigma$ (thick
solid line in left lower panel of
figure~\ref{fig:conf})\footnote{~Similar results have also been
independently obtained by Roy Maartens and Elisabetta Majerotto (private
communication).}.

We therefore conclude that while the flat Brane\,1 and Brane\,2 models
are able to satisfy the Gold and SNLS data data over a reasonable
region of parameter space, both the DGP model and the model with a
quiescent future singularity are in tension with the data.  For both
the supernova datasets, the more general flat Brane\,1 and Brane\,2
models are able to satisfy the data over a reasonable region of
parameter space.

\section{Conclusions}
\label{sec:conclusions}

We have demonstrated that high redshift type Ia supernova data
\cite{riess04,astier05} when combined with the recent discovery of
baryon oscillations in the SDSS \cite{eisenstein05} can serve to place
significant constraints on the parameter space of braneworld models.

Our results for the Gold data set are in broad agreement with the
earlier work of \cite{alam02} who tested braneworld models against an
early SNe data set (also see \cite{polish}).  
Our results for the Supernova Legacy
Survey (SNLS) are in good agreement with those of \cite{goobar} who
recently tested the DGP model using SNLS and baryon oscillations.
Since the DGP model forms a subset of the braneworld models analyzed
by us we find that the SNLS data together with baryon oscillations
rule out this model at 3 $\sigma$.  However we also find that the DGP
model is more strongly constrained by SNLS than by the Gold data set
of \cite{riess04}, which allows the DGP model at $2\sigma$. Our
analysis also shows that baryon oscillations in conjunction with SNe
data rule out a class of braneworld models in which the universe
encounters a `quiescent' future singularity where the density,
pressure and Hubble parameter remain finite but higher derivatives of
the scale factor diverge.

Our analysis indicates that the Gold and SNLS supernovae place
slightly different constraints on the braneworld parameters.  Thus
although figures~\ref{fig:chi},~\ref{fig:conf} clearly show that the
Braneworld models analyzed by us agree well with both sets of SNe
data, the Gold data set accommodates larger values of $\om \ggeq 0.25$
than SNLS ($\om \ggeq 0.2$) for the Brane\,1 model.  In the case of
Brane\,2 smaller values $\om \lleq 0.35$ appear to be favoured by SNLS
than by Gold ($\om \lleq 0.45$). Thus Brane\,1 models fit better to
the SNLS data, while Brane\,2 models fit better to the Gold dataset.

In this connection it is interesting to note that the recent analysis
of evolving dark energy models using Gold and SNLS data \cite{greek}
found somewhat different trends in the evolution of dark energy
favoured by these two data sets.  It is hoped that improvements in the
quality and quantity of future SNe data will allow tighter constraints
to be placed on dark energy models.  We would like to draw attention
to the fact that in this paper we have explored the region of
parameter space in which $\omx$ is reasonably small.  The reason for
this is that performing a likelihood analysis for the full (formally
infinite) region of parameter space is computationally very expensive
and so we have restricted our analysis to a finite region of parameter
space corresponding to $\omx \lleq 1$, $\Omega_\ell
\lleq 1$ which includes $\l$CDM and the DGP braneworld as subclasses. 
In this context it is interesting to note that braneworld models with
fundamentally new properties can arise for large values of $\omx \gg
1$. For instance, the mimicry model briefly touched upon in section
II, mimics $\Lambda$CDM at low redshifts while departing from the
latter at higher redshifts.  The presence of a `dark radiation' term
(which quantifies the projection of higher dimensional `bulk' effects
on to the brane) allows for even more radical departures from standard
cosmology by permitting the universe to `loiter' at moderately large
redshifts $1 \ll z \ll 1000$.  Both `loitering' \cite{loiter} and
`mimicry' \cite{mimic} models predict a lower redshift of reionization
and a longer age for QSO's and other high redshift objects when
compared with $\l$CDM.  However at low $z$ both models remain very
close to $\l$CDM and for this reason are unlikely to be distinguished
from the latter on the basis of SNe data alone.
%We have also not included in this analysis the possible presence of
%a `dark radiation' term which quantifies the projection of higher dimensional
%`bulk' effects on to the brane. As pointed out in \cite{loiter} the presence of
%this term can make the universe `loiter' at intermediate redshifts while 
%$w_0 \simeq -1$ at the present epoch. 
For this reason we have not included `loitering' and `mimicry' models
into the present analysis but hope to return to these models 
in a companion paper.

\section*{Acknowledgments}
We acknowledge stimulating discussions with Roy Maartens and Yuri Shtanov.
UA was supported for this work by the CSIR.  
%VS and YuS acknowledge
%support from the Indo-Ukrainian program of cooperation in science and
%technology sponsored by the Department of Science and Technology of
%India and Ministry of Education and Science of Ukraine.

\end{document}